\documentclass{article}
\usepackage[]{arxiv}

\usepackage[utf8]{inputenc} % allow utf-8 input
\usepackage{hyperref}       % hyperlinks
\usepackage{url}            % simple URL typesetting
\usepackage{booktabs}       % professional-quality tables
\usepackage{amsfonts}       % blackboard math symbols
\usepackage{nicefrac}       % compact symbols for 1/2, etc.
\usepackage{microtype}      % microtypography
\usepackage{amsmath,amssymb,amsfonts}
\usepackage{algorithmic}
\usepackage{mathrsfs}
\usepackage{graphicx}
\usepackage{textcomp}
\usepackage{xcolor}
\usepackage{dsfont}
\usepackage{amsthm}
\newtheorem{theorem}{Theorem}[]
\newtheorem{corollary}{Corollary}[]
\newtheorem{lemma}[]{Lemma}

\newtheorem*{assumption*}{Assumption}
\newtheorem{assumption}{Assumption}

\newtheorem{remark}{Remark}
\usepackage{amsmath}
\DeclareMathOperator*{\argmax}{arg\,max}
\DeclareMathOperator*{\argmin}{arg\,min}
\DeclareMathOperator*{\E}{\mathbb{E}}
\DeclareMathOperator*{\R}{\mathbb{R}}

\usepackage{mathtools}
\graphicspath{ {./images/} }
\usepackage{dsfont}
\usepackage{titling}
\usepackage[ruled,vlined]{algorithm2e}
\predate{}
\postdate{}

\date{}

\title{\LARGE \bf
Efficient Episodic Learning of Nonstationary and Unknown Zero-Sum Games Using Expert Game Ensembles
}

\author{Yunian Pan and Quanyan Zhu
\thanks{The authors are with the Department of Electrical and Computer Engineering, Tandon School of Engineering, New York University, Brooklyn, NY, 11201 USA; E-mail: {\tt\small \{yp1170,qz494\}@nyu.edu}}%
}

\begin{document}

\maketitle

%%%%%%%%%%%%%%%%%%%%%%%%%%%%%%%%%%%%%%%%%%%%%%%%%%%%%%%%%%%%%%%%%%%%%%%%%%%%%%%%
\begin{abstract}
Game theory provides essential analysis in many applications of strategic interactions. However, the question of how to construct a game model and what is its fidelity is seldom addressed. 
In this work, we consider learning in a class of repeated zero-sum games with unknown, time-varying payoff matrix, and noisy feedbacks, by making use of an ensemble of benchmark game models. 
These models can be pre-trained and collected dynamically during sequential plays. They serve as prior side information and imperfectly underpin the unknown true game model. 
We propose \texttt{OFULinMat}, an episodic learning algorithm that integrates the adaptive estimation of game models and the learning of the strategies. The proposed algorithm is shown to achieve a sublinear bound on the \textit{saddle-point regret}. We show that this algorithm is provably efficient through both theoretical analysis and numerical examples. We use a dynamic honeypot allocation game as a case study to illustrate and corroborate our results. We also discuss the relationship and highlight the difference between our framework and the classical adversarial multi-armed bandit framework.

\end{abstract}

%%%%%%%%%%%%%%%%%%%%%%%%%%%%%%%%%%%%%%%%%%%%%%%%%%%%%%%%%%%%%%%%%%%%%%%%%%%%%%%%
\section{INTRODUCTION}

Game theory has been used to model and analyze complex and strategic multi-agent interactions, and has a wide range of applications in economics, sociology, politics, and engineering. 
Game theoretic analysis often relies on the construction or estimation of the underlying game models. For example, Empirical Game Theoretical Analysis (EGTA) \cite{wellman2006methods} uses %{\bf empirical and simulated? what does it mean?}
empirical observations from a priori black-box simulator to analyze the equilibria. 
%The methodologies have been discussed in both generic and specific scenarios, e.g., online auction and cyber-security. 
%The feature in these successful EGTA research, is that we do not assume a prior access to the payoff function, but estimate it through readily existing black-box simulator. 
Model-based reinforcement learning (RL) is another approach that can guarantee the sample efficiency as well as enable learning online tasks. 
% along the way of not only finding solutions to the simulator, but also leaning real-life online tasks. 
In this work, we consider the setting where the autonomous agent aims to learn an unknown repeated zero-sum game using an ensemble of priori known  %{\bf expert?} Yes it was mentioned later
expert game models. The \textit{target game} can be either a black-box simulator, or a multi-agent task with unknown utility functions. 
% instead the knowledge of the utility functions in those benchmark models is known a priori.

%The essential difference between simulation and reality has been discussed in many control scenario, we notice that although there are means to create some nearly perfect simulators or models (e.g., by using the law of physics), the modeling error still exists. 
%To ensure accuracy it's plausible to set a variety of models as benchmarks, in doing so, one significantly reduces the inductive bias and is able to transfer previously obtained knowledge. 
%The benchmark model ensembles can be obtained from different sources using different data collecting policies, they provide more accurate approximation of the real environment.

We propose the framework of approximating an underlying unknown zero-sum game using a set of a priori zero-sum game models, which we refer to as \textit{expert games}.
%??? IS THIS BENCHMARK GAMES? YES
%, their predictions are called \textit{expert games}. 
%These expert games have the same structure as the target game, the only difference lies in their payoffs. 
%To simplify the analysis 
%This work focuses on the class of zero-sum normal-form games. % in the rest of this paper, note that analysis about general-sum dynamic games should follow similar procedure. 
In our framework, agents (players) repeatedly interact in an environment for a period of time. The total duration is divided into \textit{episodes} and every episode is divided into a fixed number of \textit{rounds}. An episode starts with revealing the expert games to the players. At each round, agents simultaneously observe an executed action pair and its corresponding noisy payoffs. The players can estimate the game based on their observations and  choose strategies that minimize the cumulative regret. 
%to make themselves unexploitable by their opponents.

%\subsection{Multi-Armed Bandits and Matrix Games}
%The agents  explore to understand the outcome of each action pair, as well as the relationship between the target game and the expert games, meanwhile, they must exploit the understood game matrix structure to design their strategy, usually in a mixed-from. Therefore, 
%We make use of both the classical Multi-Armed Bandit (MAB) \cite{lai1985asymptotically} framework and matrix game theory to design our algorithmic solution.

This work focuses on the class of zero-sum normal-form games, which can be represented by a matrix. The game consists of a row player and a column player. The game matrix is unknown to the players ahead of time.  
%They choose their actions simultaneously. %The gain of the row player is the loss of the column player. The \textit{saddle-point} equilibrium is the solution concept used to characterize the equilibrium outcome of the game. % The expected payoff achieved at the equilibrium is called the  \textit{value} of the game. 
The task of one player, say, the row player is to minimize the loss of the sequential play, no matter what her opponent does. If the game matrix were known, a security strategy would be to play the saddle-point mixed strategy at each round. One challenge of the problem is that the players need to find the best-effort strategies based on historical observations without knowing the game. 
% One shall notice that this setting can be reduced to adversarial bandit setting, as in each round the row player receives arbitrary column chosen by the column player, as discussed in \cite{cesa2006prediction}.

Multi-Armed Bandit (MAB) \cite{lai1985asymptotically}  is a fundamental sequential decision-making framework in an unknown environment.
%in RL literature, it is not only powerful to capture the exploration-exploitation behavior when facing up with uncertainty, but also fairly simple to allow for rich theoretical analysis. There have been tons of variants of MAB problems up to present, among these variants, 
The contextual bandit is a variant of the MAB framework that allows agents to make decisions with side information.  
%that arge decision sets, non-stationarity caused by ignoring side information, thus it has been applied in a broad range of real-world web service applications such as online advertising and recommendation systems \cite{li2010contextual,zheng2018drn}, and wireless communication scenarios such as network channel selection \cite{bajrachrya2021contextual} and allocation \cite{zhou2020human}. 
%{\bf THE DESCRIPTION IS COMPLICATED. NEED TO BE PARAPHRASED AND MADE EASY TO UNDERSTAND. The side information is the data in learning problems that are neither the input nor the output for the hypothesis learning object, it can be some intermediate features, inputs or outputs of the functions that are related to the hypothesis, or even relations between the inputs and outputs. }
The side information is any knowledge that is not the direct input for a learning task, but correlates with the intrinsic features of the task. 
In our framework, expert games can be viewed as the source of side information. The knowledge of expert games can be acquired in multiple ways, e.g., from the past experience of the players or a collaborative agent. 
%For example (e.g., in federated learning) or learned by different model integration methods, (e.g., injecting sample attacks into the intrusion detection system (IDS) to define utility functions.)

We develop a regret-efficient algorithm that allows the players to estimate the game using the contextual information of expert games and adapt their strategies to minimize cumulative regrets. The notion of the regret we define here is the saddle-point regret, which is slightly weaker than the best-response regret. Yet, they are equivalent when a player faces an opponent playing saddle-point strategy.
%the contextual bandit
%\subsection{Distinguishment of our Framework}
%Normally one would expect that an algorithm which solves adversarial bandit also solves the learning in games, it is however not the case in our framework. As will be presented later, 
We show that our algorithm outperforms the exponential-weight algorithm for exploration and exploitation (\texttt{Exp3}) in terms of the performance across the entire time period. This result arises from the following features of our design: 
\begin{itemize}
\item[1)] Our algorithm explores the matrix structure and takes advantage of the knowledge of the expert games; % \texttt{Exp3} does not explore the entire matrix structure, nor does it take advantage of the knowledge of expert games; %while our algorithm utilizes the expert predictions; 
\item[2)] Our algorithm achieves a near-optimal policy that is sufficient to achieve the values under the saddle-point equilibria. %\texttt{Exp3} does not guarantee convergence to the equilibrium, while our algorithm achieves a near-optimal policy;
\item[3)] Our algorithm is aware of the time-varying nature of underlying matrix game in contrast to \texttt{Exp3}. 
% as the game matrix changes over time, and so is the hindsight optimal action.
\end{itemize}
%A formal treatment will be presented in later section.

%\subsection{Applications of Zero-sum Game}
%Applications of zero-sum games may include but are not limited to: 1. robust control, where the plant is vulnerable to adversarial control injection; 2. financial market, where the trading is sometimes viewed as a zero-sum game with winners and losers. 3. political election, where targeting message at different group of voters might have different impact for competing parties. 
%WHAT ARE THE RESULTS OF THE ALGORITHM???
A direct result of the algorithm is that the agent can suffer only $\tilde{\mathcal{O}}(\sqrt{KT})$ saddle-point regret in the sequential play (here $K$ is the total number of episodes and $T$ is the total number of rounds in each episode), while theoretical result shows that \texttt{Exp3} can only reach a $\tilde{\mathcal{O}}(K\sqrt{T})$ bound on best-response regret. We also show that the agent can quickly learn the weighting coefficients of the expert games, as the interaction proceeds.

One important application of the proposed algorithm is in the domain of cybersecurity. The interaction between an attacker and a defender naturally leads to zero-sum game scenario. However, in many cases, the knowledge of the true underlying game is expensive and non-stationary. For example, in network systems, the exact configuration information for every individual node may be incomplete, and the network states would often be time-varying. %The network systems would rely on techniques, such as EGTA and RL, to evaluate the true attack model, and learn the equilibria from a black box simulator. These techniques finally result in approximate payoff functions that can be used as model ensembles. 
Our framework can leverage the expert domain knowledge of the network and form an ensemble of expert security games. They represent the expert knowledge of the attack model as of the episode the game is played. The network system may not encounter the same attacker in each episode. The underlying security game can change over the episodes. Furthermore, a sophisticated attacker may exploit a zero-day vulnerability unknown to the defender. By the end of the episode, the defender will learn about the zero-day attacker and expand the expert domain knowledge by including the experienced zero-day attack encoded as a new expert security game.
The proposed algorithm will enable the learning and adaptation of the defense strategies at each round and episode to secure the network. %seek the stationary pattern between these learned models and the true model, and enable autonomous agents to make secure online decisions. 

We present our problem formulation along with discussions in Section \ref{problemformulation}.  Related works will be discussed in Section \ref{relatedwork}. The algorithmic development and theoretic analysis will be presented in Section \ref{theoreticalanalysis}. A case study and conclusions will be stated in Sections \ref{casestudy} and \ref{conclusion}, respectively.

%DRAW A TIME LINE TO ILLUSTRATE THE PLAY OF THE GAME. SPECIFY THE INFO

%MORE EVALUATIONS AND MAKE SOME CONCLUSIONS FROM CASE STUDIES.

\section{PROBLEM FORMULATION}\label{problemformulation}

\subsection{Non-stationary Games with Side Information}

\begin{figure}
    \centering
    \includegraphics[width = .9\textwidth]{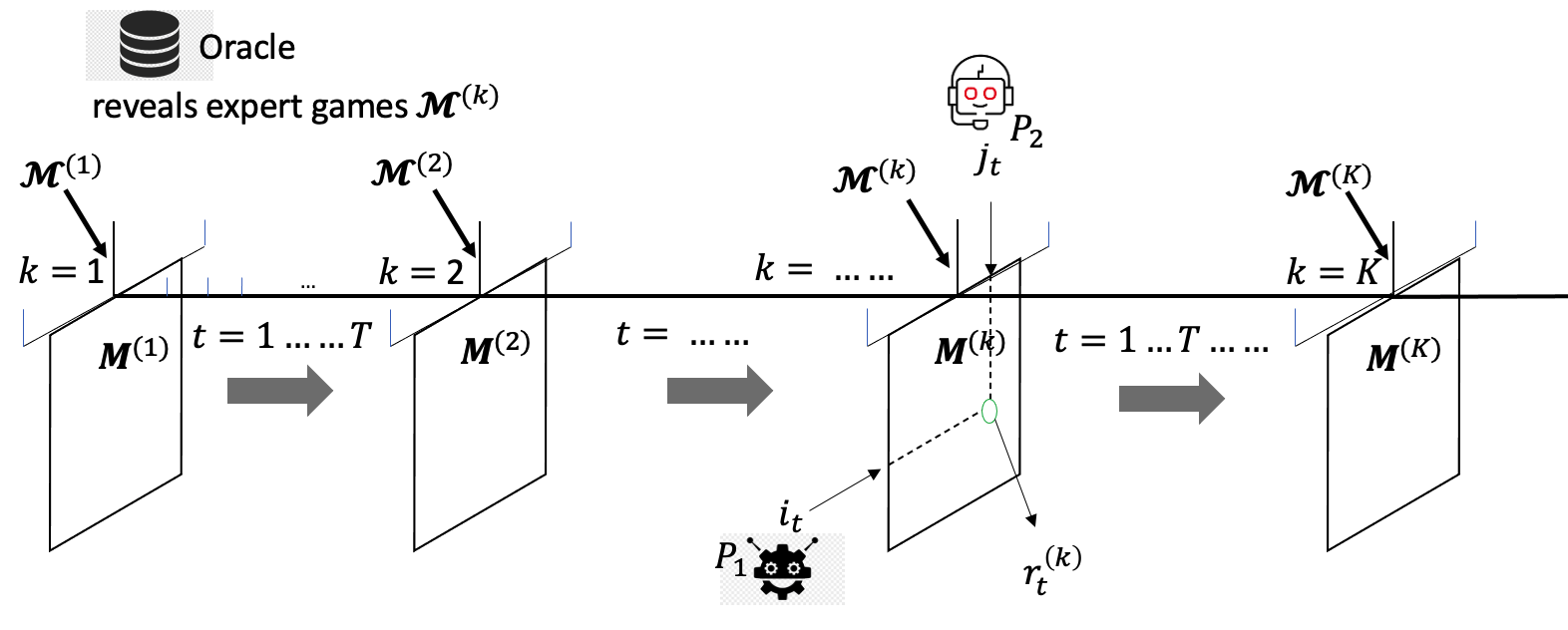}
    \caption{Illustration of the timeline of episodic learning. The learning is divided into multiple episode. Each episode has multiple rounds. Oracle reveals the expert game $\boldsymbol{\mathcal{M}}^{(k)}$ at the beginning of each episode. At round $t$ in an episode, the players choose the action pair $(i_t, j_t)$ of an unknown zero-sum matrix game $\boldsymbol{M}^{(k)}$. The payoff entry $\boldsymbol{M}^{(k)}_{i_t, j_t}$ is observed by the players at each round. The underlying unknown game $\boldsymbol{M}^{(k)}$ may change at each episode $k$.}
    \label{timeline}
\end{figure}

We define a general class of repeated two-player zero-sum game containing multiple time scales and expert information. It can be encapsulated by the tuple 
$\big\langle \{\mathrm{P}_1, \mathrm{P}_2\}, \{[n_1], [n_2] \}, K, T, \{\boldsymbol{M}^{(k)}, \boldsymbol{\mathcal{M}}^{(k)}\}_{k=1}^K \big\rangle$.
This game is played by a row player $\mathrm{P}_1$ (the maximizer) and a column player $\mathrm{P}_2$ (the minimizer). 
Their action sets are $[n_1] = \{1 ,\ldots, n_1\}$ and $[n_2] := \{1,\ldots, n_2\}$, respectively, and remain unchanged during the play. 
The duration of their interactions is divided into episodes $k = 1, \ldots, K$, and each episode contains a finite number of time fractions $t = 1, \ldots, T$. 
The payoff matrix $\boldsymbol{M}^{(k)} \in \R^{n_1 \times n_2}$ defines the target game or \textit{ground truth game} underlying the environment for each episode $k$. It is naturally evolving across episodes, but at the finer time scale $t$, it is time-invariant, so that in each episode players play a static game for finite rounds.
Associated with each $\boldsymbol{M}^{(k)}$ is a set of matrices that define the expert games, they are also called side information or \textit{contextual matrices}, and are revealed to the players at the beginning of every episode. The expert games $\boldsymbol{\mathcal{M}}^{(k)} = \{ M^{(k)}_1, \ldots, M^{(k)}_S \}$, where the cardinality $|\boldsymbol{\mathcal{M}}^{(k)}| = S$. Let subscripts $s =1,\ldots, S$ index
expert games. We impose a regularity assumption that each $M^{(k)}_s \in [0,1]^{n_1 \times n_2}$ is of the same dimension as the target game and they capture the same action capability of both players. Figure \ref{timeline} illustrates the timeline of the sequential play.
%INDEX K SHOULD BE CAPITALIZED IN THE FIGURE? MORE INFORMATION IN THE FIGURE? 

Denote the strategies of $\mathrm{P}_1$ and $\mathrm{P}_2$ by $\mu \in \Pi^1 $, $\nu \in \Pi^2 $, with their strategy space $\Pi^1$ and $\Pi^2$ being $(n_1 - 1)-$dimensional and $(n_2 - 1)-$dimensional simplex, respectively, i.e., $\Pi^1 : = \Delta(n_1)$ and $\Pi^2 := \Delta(n_2)$. 
In every episode $k$,  there exists a mixed-strategy \textit{saddle-point} equilibrium that is optimal for both players. We denote this mixed-strategy saddle-point value of the ground truth game $\boldsymbol{M}^{(k)}$ as $\mathrm{val}(\boldsymbol{M}^{(k)})$:
$$
 \mathrm{val}(\boldsymbol{M}^{(k)}) : = \sup_{\mu \in \Pi^1} \inf_{\nu \in \Pi^2 } \mu^{\top} \boldsymbol{M}^{(k)} \nu  = \inf_{\nu \in \Pi^2 } \sup_{\mu \in \Pi^1} \mu^{\top} \boldsymbol{M}^{(k)} \nu 
$$

The players' goal is to play as close to the saddle-point strategy as possible to ensure robustness in a minimax sense. Yet they have no knowledge about $\boldsymbol{M}^{(k)}$. Hence the players need to exploit the expert information to estimate the game using their accumulated observations.

\subsection{Parametric Assumption and Learning Protocol}
\subsubsection{Linear Combinations of Expert Games}
We aim to find the implications of expert predictions by exploring the stationary pattern, this pattern is encoded by a parametric assumption relating the underlying payoff matrix $\boldsymbol{M}^{(k)}$ and these contextual matrices, which states the following:
\begin{equation*}
\boldsymbol{M}^{(k)} = \boldsymbol{h}(\boldsymbol{\mathcal{M}}^{(k)}; \theta^*) 
\end{equation*}
where the matrix-valued function $\boldsymbol{h}: [0,1]^{S \times n_1 \times n_2} \to \R^{n_1 \times n_2}$ is assumed to be stationarily parameterized by $\theta^*$ living in some parameter space. 
In particular, we assume that $\boldsymbol{h}$ takes a linear form, i.e., $\theta^* \in \R^S$ and %{\bf THE ESTIMATED GAME IS THE LINEAR COMBINATION. THE TRUE GAME MIGHT NOT BE the ground truth game is a linear combination of expert games.}
the estimated ground truth game is a linear combination of expert games.
Therefore, given a set of expert games, $\boldsymbol{h}$ can be expressed as a function of the variable $\theta$:
\begin{equation} \label{linearassumption}
     \boldsymbol{h}( \boldsymbol{\mathcal{M}}^{(k)}; \theta) = \sum_{s=1}^S \theta_s M^{(k)}_{ s } 
\end{equation}

%A similar example is seen in RL literature \cite{DBLP:journals/corr/abs-1910-10597}, which uses state-action dependent decomposition.

\subsubsection{Learning through Feedback}
While being aware of the contextual matrices, the players do not have access to the true game; instead they learn the underlying matrix $\boldsymbol{M}^{(k)}$ through sequential interactions: at each round $t$ of episode $k$, players choose their own actions $i_t \in \{1, \ldots, n_1\}$ and $j_t \in \{1,\ldots, n_2\}$, and obtain their payoffs $r^{(k)}_t$ and $-r^{(k)}_t$:
\begin{equation}\label{rewardfeedback}
    r^{(k)}_t = \boldsymbol{M}^{(k)}_{[i_t, j_t]} + \eta_t^{(k)}
\end{equation}
Here, we use $\boldsymbol{M}^{(k)}_{[i,j]}$ to denote the $i_t,j_t$-th entry of the matrix $\boldsymbol{M}^{(k)}$.
The noise $\eta_t^{(k)}$ is assumed to be a Martingale difference sequence that is conditionally $1$-sub-Gaussian, i.e.,
\begin{equation*}
    \begin{aligned}
      \E [ \eta_t^{(k)}  | \mathcal{H}^{(k)}_{t-1} ]  = 0, \quad
      \E(\exp(\gamma\eta_t^{(k)} ) | \mathcal{H}^{(k)}_{t-1})  \leq \exp(\frac{\gamma^2}{2}),
    \end{aligned}
\end{equation*}
where $\mathcal{H}^{(k)}_{t-1} := \{ (i^{(k')}_t, j^{(k')}_t, r^{(k')}_t)_{t=1}^T , \boldsymbol{\mathcal{M}}^{(k')} \}_{k'=1}^{k-1}$ $\bigcup$ $\{(i^{(k)}_t, j^{(k)}_t, r^{(k)}_t)_{t=1}^{t-1}, \boldsymbol{\mathcal{M}}^{(k)}\}$ is the historical observations up to episode $k$ and prior to time $t$. 
We introduce $\sigma(\mathcal{H}^{(k)}_{t-1})$ as the $\sigma-$field generated by the history. 
 Without loss of generality, we only discuss the control of $\operatorname{P}_1$. Thus, at time $(k-1)T + t$, an algorithm \texttt{alg} is a $\sigma(\mathcal{H}^{(k)}_{t-1})$-measurable mapping from the set of possible histories $\mathcal{H}^{(k)}_{t-1}$ to the strategy space $\Pi^1$, generating strategy $\mu^{(k)}_t$. We denote $\operatorname{P}_2$'s strategy as $\nu^{(k)}_t$, but we do not make specific assumptions about $\nu^{(k)}_t$. % whether $\operatorname{P}_2$ is a learning agent or not is beyond the scope of this paper.

\subsection{Performance Metric}
A plausible goal of learning in such an environment from one player's perspective is to minimize the cumulative regret against some benchmark strategies. In a single-agent setting, the benchmark strategy is the hindsight optimal action, but since we have a multi-agent environment, the hindsight optimum is changing round by round, this makes tracking the internal regret per round hard to analyze. Here we discuss several cases of regret definition from one player's perspective. 

\subsubsection{Best Response Regret}
The \textit{best-response regret} is defined as the gap between the actual expected performance and the expected outcome of best response against the opponent's mixed-strategy. Let $\nu_t^{(k)}$ be the mixed-strategy that $\mathrm{P}_2$ uses in episode $k$ at round $t$, player $\mathrm{P}_1$'s best response regret $B\mathcal{R}$ of $\texttt{alg}$ during the $K$ episodes play is defined as:
\begin{equation*}
    B\mathcal{R}( \operatorname{alg}, K) =  \mathbb{E}_{\eta, \texttt{alg}}\left[ \sum_{k= 1}^{K}\sum_{t=1}^T \sup_{ \mu \in \Pi^1} \mu^{\top}\boldsymbol{M}^{(k)}\nu_t^{(k)} - r^{(k)}_t\right]
    \label{regret}
\end{equation*}
where the randomness comes from the learner's algorithm as well as the noise.

%The best-response regret generalizes the internal regret per round. If we restrict $\mu^{(k)}_t$ and $\nu^{(k)}_t$ to be pure strategies, its definition coincides with internal regret, which tracks the best hindsight action per round. However, bounding this regret is difficult since we do not have knowledge about $\nu^{(k)}_t$. 

\subsubsection{Exploitability Regret}
The \textit{exploitability regret} is defined by adding up two player's best-response regret. 
%It is usually useful when two agents use the same learning algorithm, namely, {\bf IT SEEMS NOT FORMAL doing self-play. }

\begin{equation*}
  \begin{aligned}
      E\mathcal{R}(\texttt{alg}, K) & = \mathbb{E}_{\eta, \texttt{alg}}\bigg[\sum_{k= 1}^{K}\sum_{t=1}^T \sup_{ \mu \in \Pi^1} \mu^{\top}\boldsymbol{M}^{(k)}\nu_t^{(k)} - \\  & \quad \quad \quad \quad \quad \quad 
      \inf_{\nu \in \Pi^2} \mu_t^{(k)\top}\boldsymbol{M}^{(k)}\nu \bigg] \\
      & = \underbrace{\mathbb{E}_{\eta, \texttt{alg}}\left[\sum_{k= 1}^{K}\sum_{t=1}^T \sup_{ \mu \in \Pi^1} \mu^{\top}\boldsymbol{M}^{(k)}\nu_t^{(k)} - r^{(k)}_t \right]}_{B\mathcal{R}\ \ of\ \ \mathrm{P}_1}  \\ &   + \underbrace{\mathbb{E}_{\eta, \texttt{alg}}\left[ \sum_{k= 1}^{K}\sum_{t=1}^T r^{(k)}_t - \inf_{\nu \in \Pi^2} \mu_t^{(k)\top}\boldsymbol{M}^{(k)}\nu \right]}_{B\mathcal{R}\ \ of\ \ \mathrm{P}_2}
  \end{aligned}
\end{equation*}
If the exploitability regret is asymptotically small, the two-agent system will closely reach the saddle-point equilibrium. However, oftentimes the exploitability regret analysis is limited to self-play or scenarios where both player's learning procedures are known.

\subsubsection{Saddle-Point Regret}
We focus on \textit{Saddle-Point regret}, which is defined using the gap between the saddle-point value and the actual expected performance. The saddle-point regret $S\mathcal{R}$ for $\mathrm{P}_1$ using $\texttt{alg}$ during $K$ episodes of play is:
\begin{equation}
    S\mathcal{R}(\texttt{alg}, K) =  \mathbb{E}_{\eta, \texttt{alg}}\left[ \sum_{k= 1}^{K} \sum_{t= 1}^{T} \operatorname{val}(\boldsymbol{M}^{(k)}) -  r^{(k)}_t\right].
     \label{saddlepointregret}
\end{equation}
We define the \textit{pseudo saddle-point regret} $S\hat{\mathcal{R}}$ as a random difference between the saddle-point value and the expected reward:
\begin{equation*}
    S\hat{\mathcal{R}}(\texttt{alg}, K) = \sum_{k= 1}^{K} \sum_{t= 1}^{T} \operatorname{val}(\boldsymbol{M}^{(k)}) -  \mu^{(k)}_t \boldsymbol{M}^{(k)} \nu^{(k)}_t.
 \end{equation*}

\subsubsection{Discussion and Comparison with Adversarial Setting}
For a fixed sequence of strategies executed:
\begin{equation*}
    S\mathcal{R}   \leq  B\mathcal{R} \leq    E\mathcal{R}.
\end{equation*}
Therefore, guaranteeing small $E\mathcal{R}$ or $B\mathcal{R}$ naturally guarantees small $S\mathcal{R}$. However, on the one hand, small $E\mathcal{R}$ requires the regret of opponent being small. On the other hand, $S\mathcal{R}$ analysis requires tracking the opponent's strategy, thus further assumption about the opponent is usually needed. 
We note that since in many cases the opponent might not even be a learning agent, it is ideal to ensure good performance regardless of what the opponent's strategy is, which motivates us to define the saddle-point regret. 

It is also worth noting that our framework can be viewed as an adversarial MAB problem, by rewriting \eqref{rewardfeedback} as $r^{(k)}_t = \boldsymbol{M}^{(k)}_{t, i_t}$, where $ \boldsymbol{M}^{(k)} \in \R^{T \times n_1}$ is the sequence of column chosen by the opponent plus the random noise. Thus, the \texttt{Exp3} algorithm is applicable to our problem. The details of the algorithm can be found in \cite{cesa2006prediction}. %It's questionable whether adversarial treatment is a better solution. 
%Well it's not. 
Adversarial MAB algorithms such as \texttt{Exp3} and their variants usually deal with the external regret $\mathcal{R}$. When applied to a static matrix game, the regret is defined as following:
\begin{equation}\label{externalregret}
    \mathcal{R}(\texttt{alg}, K) =  \mathbb{E}_{\eta, \texttt{alg}}\left[ \sum_{k= 1}^{K} \max_{i \in [n_1]}\sum_{t= 1}^{T} \boldsymbol{M}^{(k)}_{i, j_t} -  r^{(k)}_t\right]
\end{equation}
This regret is slightly larger than saddle-point regret: $\mathcal{R} \geq S\mathcal{R}$. However, applying \texttt{Exp3}-type of algorithms for $K$ times will cause the agent to suffer linear regret, while our algorithm reaches a bound that is sublinear in $K$. 
%{\bf I DO NOT UNDERSTAND THIS SENTENCE, NEED TO BE PARAPHRASED WHO IS OPERATOR? If the algorithm is not reset across episodes, the $\max_{i \in [n_1]}$ operator shall be moved out of $\sum_{k= 1}^{K}$, } 
Or, applying \texttt{Exp3} for once would only ensure the regret comparing to a hindsight optimal action (i.e., the regret defined as $\mathbb{E}_{\eta, \texttt{alg}}\left[ \max_{i \in [n_1]}\sum_{k= 1}^{K} \sum_{t= 1}^{T} \boldsymbol{M}^{(k)}_{i, j_t} -  r^{(k)}_t\right]$) to be small, this performance metric is weaker than what is defined in \eqref{saddlepointregret}. 

%%%%%%%%%%%%%%%%%%%%%%%%%%%%%%%%%%%%%%%%%%%%%%%%%%%%%%%%%%%%%%%%%%%%%%%%%%%%%%%%
\section{RELATED WORK}\label{relatedwork}

\subsection{Learning with a Mixture of Models}
Using multiple game models to approximate the real environment was proposed in \cite{pan2020masage}, it assumes that the existence of equilibrium data and focuses on the parametric identification and numerical solution. However, this idea of model mixtures can be traced back to multiple model RL \cite{doya2002multiple}.  The authors decompose the given task domain into a convex combination of multiple models. Instead of estimating the coefficients, they aim to train the model ensembles and their mixture weights. The linear combination framework was proposed in \cite{DBLP:journals/corr/abs-1910-10597}, which decomposes the reward and transition function using state-action dependent features. The model uncertainty setting receives wide interests and is studied in %{\bf WHAT IS SINGLE? } 
single-agent RL literature, its natural extension to the multi-agent setting, however, is rarely studied.

\subsection{Robust Contextual Multi-Armed Bandits}
A game with noisy payoffs can be viewed as a special case of robust MAB problems, see \cite{o2020matrixgamebanditfeedback,caro2015robust}. Because from one player's perspective, the actions can be viewed as arms, but the outcome of each arm is partially determined by another player's action.
Our framework can be viewed as a contextual extension of it, which is a special case of the robust MAB problem. Since the game model ensembles can be viewed as context for each action/arm.
Nevertheless, unlike general robust MAB problems, the capability of the adversary is encoded into a matrix, which makes its action more predictable.
Thus, equipped with a game theoretic viewpoint, the robust strategy is to make the worst-case best-response in a minimax sense.

\subsection{Learning in Games with Side Information}
Another viewing angle is to treat this problem as learning in games with side information, a closely related formulation has been studied in \cite{sessa2020contextual}. In studying the repeatedly played game driven by context information, the authors have modeled the correlation between game payoff and contexts using kernel-based non-parametric methods. However, the focus of their work is to study convergence to an newly proposed equilibria concept, called contextual coarse correlated equilibria, and its efficiency in a more generic setting. 
Our work targets zero-sum cases with model ensembles representing the context, this additional information enables parametric estimation and %{\bf NOT CLEAR WHAT PLANNING MEANS, PLANNING OF WHAT? allows for direct planning.}
allows for direct computation of saddle-point strategies.

%%%%%%%%%%%%%%%%%%%%%%%%%%%%%%%%%%%%%%%%%%%%%%%%%%%%%%%%%%%%%%%%%%%%%%%%%%%%%%%%
\section{THEORETICAL ANALYSIS}\label{theoreticalanalysis}

The key idea for ensuring learning efficiency is to implement the optimistic exploration of underlying game entries and select actions accordingly. This principle is called \textit{Optimism in the Face of Uncertainty} (OFU). Note that since \eqref{linearassumption} holds, exploring one game entry will give information about other game entries, as all the action-payoffs are correlated through the parameter $\theta^*$.
Thus, estimating the true parameter $\theta^*$ is essential to the algorithmic development. 
To do this, we leverage a Kalman-filter-type result to obtain an adaptive minimum-variance estimation $\hat{\theta}^{(k)}$. We construct the confidence region around $\hat{\theta}^{(k)}$ for optimistic exploration of the underlying matrix $\boldsymbol{M}^{(k)}$.

\subsection{Online Confidence Set Construction}

At round $t$ of episode $k$ , $\operatorname{P}_1$ and $\operatorname{P}_2$ jointly choose the entry of the unknown game matrix that corresponds to row $i_t$ and column $j_t$, and the payoff of this entry is partially revealed by the expert games.
Let 
$$z^{(k)}_{i_t, j_t} := \left([M^{(k)}_1]_{i_t, j_t}, [M^{(k)}_2]_{i_t, j_t}, \ldots,  [M^{(k)}_S]_{i_t, j_t}\right)
$$
be the vector that contains the side information of one game entry, then the payoff of $\operatorname{P}_1$ playing $i_t$ and $\operatorname{P}_2$ playing $j_t$ is:
\begin{equation}
    r^{(k)}_t = \langle \theta^*,  z^{(k)}_t\rangle + \eta_t^{(k)}, 
    \label{linearmodel}
\end{equation}
where we use the shorthand notation $z^{(k)}_t$ for $z^{(k)}_{i_t, j_t}$.
We employ a minimum-variance estimation framework regularized by $l_2$-norm of $\theta$,
\begin{equation}
    \min_{\theta} \sum_{k'= 1}^{k-1} \sum_{t= 1}^T (r^{(k')}_t - \langle \theta, z_{t}^{(k')} \rangle)^2 + \lambda \| \theta\|^2_2
\end{equation}
Thus, the $\lambda$-$l_2$-regularized least square estimator of $\theta^*$ is $\hat{\theta}^{k}$:
\begin{equation}\label{episodicestimate}
\begin{aligned}
     \hat{\theta}^{(k)} &  = \left(\lambda I + \sum_{k'=1}^{k-1} \sum_{t= 1}^{T} z^{(k')}_t z^{(k')\top}_t\right)^{-1} \sum_{k'=1}^{k-1} \sum_{t= 1}^{T}z^{(k')}_t r^{(k')}_t  \\ 
      & =  V_{k-1}^{-1} Y_{k-1},
     \end{aligned}
\end{equation}
where the normalizing matrices $V_k$ is defined in \eqref{matrixcovariates}, in which we introduce its imaginary intermediate form $V_{k-1,t}$,
\begin{equation}\label{matrixcovariates}
\begin{aligned}
     V_{k-1} & := \lambda I + \sum_{k'=1}^{k} \sum_{t= 1}^{T} z^{(k')}_t z^{(k')\top}_t \\
    V_{k-1, t} & := \lambda I + \sum_{k' < k} \sum_{t'= 1}^T z^{(k')}_{t'}z^{(k')\top}_{t'} + \sum_{t'= 1}^t z^{(k)}_{t'}z^{(k)\top}_{t'}, 
\end{aligned}
\end{equation}
where $k = 1, \ldots$, and the summation of $z^{(k)}_t r^{(k)}_t$,
$$
Y_k : =  \sum_{k'=1}^{k} \sum_{t= 1}^{T}z^{(k')}_t r^{(k')}_t
$$

Since the noise $\eta_t^{(k)}$ is $1$-sub-Gaussian, we have an elliptical $\delta$-confidence ball $\mathcal{C}_k(\delta)$:
\begin{equation}\label{confidenceball}
    \mathcal{C}_{k}(\delta)=\left\{\theta \in \mathbb{R}^{S}: \|\theta-\hat{\theta}^{(k)}\|^2_{V_{k-1}} \leq \beta_{k}(\delta)\right\}
\end{equation}
where $\delta > 0$ is a confidence parameter and the ball radius term
\begin{equation}\label{betaradius}
    \beta_{k}(\delta)=\left( \sqrt{2 \ln \left(\frac{\operatorname{det}\left(V_{k}\right)^{1 / 2}
\operatorname{det}(\lambda I)^{-1 / 2}}{\delta}\right)}+\lambda^{1 / 2} B\right)^{2}
\end{equation}

Before we introduce a technical lemma that is central to our analysis of the algorithm, it is useful to impose a regularity assumption over the parameter space.
\begin{assumption}
 The true underlying parameter $\theta^*$ is inside a Euclidean ball $\mathcal{B}$ with probability $1$, where $\mathcal{B}$ is defined as:
 \begin{equation*}
     \mathcal{B} = \left\{\theta \in \mathbb{R}^S:  \|\theta\|^2 \leq B^2\right\}.
 \end{equation*}
 \label{trueinaball}
\end{assumption} 
Now we arrive at the following lemma which allows us to quantify the probability of constructed confidence bounds.
\begin{lemma}[\cite{abbasiyadkori2011online} Corollary 10.]
  With hidden linear model \eqref{linearmodel} satisfied by $\theta^* \in \R^d$, and assumption \ref{trueinaball} is given, consider the $\lambda$-$l_2$-regularized least square estimation $\hat{\theta}$ defined in \eqref{episodicestimate} with $\lambda > 0$, where $z^{(k)}_t$ are arbitrary random sequences; $V_k$ are designed as in \eqref{matrixcovariates} to indicate the inverse of the covariance matrix. Then, with probability at least $1 - \delta$, 
  \begin{equation}
   \forall \  k > 0  \quad\quad  \| \theta^* - \hat{\theta}^{(k)} \|^2_{V_{k-1}} \leq \beta_k(\delta),
  \end{equation}
  where $\beta_k(\delta)$, as defined in \eqref{betaradius}, is an increasing sequence.
  \label{lemmaofconfidenceball}
\end{lemma}

Equivalently, define $E_{k}$ as the event such that $ \theta^* \in \mathcal{C}_k(\delta)$ holds, where the elliptical ball $\mathcal{C}_{k}(\delta)$ is defined in \eqref{confidenceball}. 
Then, we have
$\mathbb{P}\left(\exists k \in \mathbb{N}^+: {E_{k}}^c\right) \leq \delta$.

\subsection{The Upper Confidence Bound Algorithm}

Our designed procedure, called Optimism in the Face of Uncertainty for Linear Matrix (\texttt{OFULinMat}), is shown in Algorithm \ref{frequentistalgorithm}.
We clarify that, at every episode $k$, the row player is executing a fixed policy $\mu^{(k)}$ at the smaller time scale $t \in [T]$. The reason is due to economic considerations. On one hand, it is not computationally efficient to update the estimation of $\theta$ whenever encountering a new reward feedback. On the other hand, since the underlying game models are consistent within the smaller time scale, it is not necessary to adjust the policy per time step.

\begin{algorithm}[htbp]
\label{frequentistalgorithm}
\SetKwInOut{Input}{Input}
\SetAlgoLined
\Input{ $B > 0$,  $0 < \delta < 1 $, $\lambda > 0$; }
Initialize $V_0 = \lambda I$, and $Y_0 = 0$\;
\For{$k = 1, 2, \ldots$}{
    Oracle reveals $M^{(k)}_1, M^{(k)}_2, \ldots, M^{(k)}_S$ \;
    Estimate $\hat{\theta}^{(k)} = V_{k-1}^{-1}Y_{k-1}$ \;
    Compute $\tilde{\theta}^{(k)} \in \argmax_{\theta \in \mathcal{C}_k(\delta) \bigcap \mathcal{B}} \operatorname{val}(\sum_{s=1}^S \theta_s M_s^{(k)}) $  \;
     Compute $\mu^{(k)} \in \argmax_{\mu \in \Pi^1} \min_{\nu \in \Pi^2} \mu^{\top} (\sum_{s=1}^S \tilde{\theta}^{(k)}_s M_s^{(k)}) \nu $ \;
   \For{$t = 1$ \KwTo $T$}{
   Execute $i_t^{(k)} \sim \mu^{(k)}(\cdot)$ \;
   %$v_t = K^2_t(\tilde{\Theta}^{\texttt{lo}}_k) x_t$ \;
   Receive opponent's action $j_t$, and reward $r_t$ \;
   }
   Save $(z^{(k)}_1, z^{(k)}_2,  \ldots, z^{(k)}_T)$ and $(r^{(k)}_1, r^{(k)}_2 \ldots, r^{(k)}_T)$ as $Z_k^{\top}$ and $X_k^{\top}$\; 
   $V_{k} := V_{k-1} + Z^{\top}_k Z_k$\;
   $Y_{k} := Y_{k-1} + Z^{\top}_k X_k$\;
}
\caption{Optimism in the Face of Uncertainty for Linear Matrix (\texttt{OFULinMat})}
\end{algorithm}

\subsection{Main Results}
\begin{assumption}
[Regularity Assumption]~
\begin{enumerate}
    \item[(a)]  For any opponent strategy $\nu$, the maximum regret gap is:
    \begin{equation*}
        \max_{k \in [K]} \sup_{\mu_1, \mu_2 \in \Pi^1} (\mu_1 - \mu_2)^{\top} \boldsymbol{M}^{(k)} \nu \leq 1.
    \end{equation*}
    \item[(b)]  The increment ratio of $\operatorname{det}(V_{k})$ is bounded by constant $\kappa -1$ ($\kappa > 1$) at every episode, i.e., 
    \begin{equation}
        \frac{\operatorname{det}(V_{k})}{\operatorname{det}(V_{k-1})} \leq  \kappa, \quad\quad k = 1, 2, \cdots
    \end{equation}
\end{enumerate}
\label{regularity}
\end{assumption}

\begin{remark}~
\begin{enumerate}
    \item[(a)]  The first assumption restricts the regret per time step to be limited by a constant $1$ so we do not have to repeat it in later analysis.
    \item[(b)]  The assumption that limits the increment ratio of determinant is reasonable since we can write $\frac{\det V_{k,t} }{\det V_{k,t-1}} = 1 + \|z^{(k)}_t z^{(k){\top}}_t\|_{V^{-1}_{k,t-1}}$, while $\|z^{(k)}_t\|$ does not grow over time, $V_{k,t-1}$ are constantly increasing, thus we can safely assume that $\det V_k$ does not change too much over the limited $T$ time-step duration.
\end{enumerate}
\end{remark}

\begin{theorem} \label{pseudoregretofulinmat}
 Under the conditions of assumptions \ref{trueinaball} and \ref{regularity}, with probability $1-\delta$, the pseudo saddle-point regret of \texttt{OFULinMat} with regularizing parameter $\lambda$ satisfies
 \begin{equation}
      S\hat{\mathcal{R}}(\texttt{OFULinMat}, K) \leq \sqrt{ 8 \kappa KTS \beta_K(\delta) \ln\left(\frac{\lambda S + KTS}{\lambda S}\right)},
 \end{equation}
 where $\kappa$ is a sufficiently large constant, and $\sqrt{\beta_K(\delta)}$ is chosen to be
 \begin{equation*}
      \sqrt{\beta_K(\delta)} = \sqrt{\lambda}B + \sqrt{2 \ln(\frac{1}{\delta}) + S\ln \left( \frac{\lambda S +  KTS}{\lambda S}\right) }.
 \end{equation*}
\end{theorem}

A direct result obtained from Theorem \ref{pseudoregretofulinmat} is that the expected regret of algorithm \texttt{OFULinMat} can be bounded by 
\begin{corollary} \label{corollary}
 Under the conditions of assumptions \ref{trueinaball} and \ref{regularity}, choosing $\delta = {1}/{KT}$, the saddle-point regret of \texttt{OFULinMat} with regularizing parameter $\lambda$ satisfies:
 \begin{equation}
      S\mathcal{R}(\texttt{OFULinMat}, K) \leq \tilde{\mathcal{O}}(S\sqrt{KT})
 \end{equation}
\end{corollary}

\subsection{ Regret Analysis } 
\subsubsection{Technical Lemmas}.

\begin{lemma}[Lemma 15, \cite{abbasiyadkori2011online}]
  Let $A, B \in \R^{S \times S}$ be two positive semi-definite matrices such that $A \succ B$, then following identity holds:
  \begin{equation*}
       \sup_{z \in \R^S, \ z \neq 0} \frac{z^{\top} A z }{z^{\top} B z } \leq \frac{\operatorname{det}(A)}{\operatorname{det}(B)}.
   \end{equation*}
   \label{technicaltool1}
\end{lemma}

\begin{lemma}[Bounding $\|z_{t}^{(k)}\|^2_{V_{k-1}^{-1}}$]
 Let $V_0 \in \R^{S\times S}$ be positive definite and the sequence of vectors $z_t^{(k)} \in \R^S$ satisfies $\|z_t^{(k)}\|_2 \leq \sqrt{S} < +\infty$ for all $t \in [T]$ and $k \in [K]$, let $V_{k-1, t}$ be as defined in \eqref{matrixcovariates}. Then,
 \begin{equation*}
 \begin{aligned}
        \sum_{k=1}^K \sum_{t = 1}^T \min \{ 1, \| z_{t}^{(k)} \|^2_{V_{k-1, t}^{-1}} \} &\leq 2\ln\left(\frac{\operatorname{det}(V_{K})}{\operatorname{det}(V_{0})}\right) \\
        &\leq 2S\ln \left(\frac{\operatorname{tr}(V_0) + KTS}{S\operatorname{tr}(V_0)^{\frac{1}{S}}}\right),
 \end{aligned}
 \end{equation*}
 where $\operatorname{tr}(\cdot)$ is the trace operator of a matrix. 
 \label{boundingz}
\end{lemma}
The result can be obtained by extending \cite{abbasiyadkori2011online} lemma 4.

\subsubsection{The Proof of Main Results}

Now we are able to establish the saddle-point regret bound via the technical tools provided above.

Let $\widetilde{\boldsymbol{M}}^{(k)}$ be the overestimation of matrix game $\boldsymbol{M}^{(k)}$, $\nu_*^{(k)} := \argmin_{\nu \in \Pi^2} \max_{\mu \in \Pi^1} \mu^{\top} \widetilde{\boldsymbol{M}}^{(k)} \nu $ be the saddle-point strategy of $\operatorname{P}_2$.
By definition, the algorithm returns a corresponding optimal strategy $\mu^{(k)}$ for $\operatorname{P}_1$ at every episodes. In a word, we have strategy profiles: 
$$
(\mu^{(k)}, \nu^{(k)}_*) = \arg \max_{\mu \in \Pi^1} \min_{\nu \in \Pi^2}\mu^{\top} \widetilde{\boldsymbol{M}}^{(k)} \nu.
$$

By the definition of the saddle-point regret:
\begin{equation*}
     \begin{aligned}
    S \mathcal{R} & = \E\left\{ \sum_{k=1}^K T\operatorname{val}(\boldsymbol{M}^{(k)})  - \sum_{t=1}^{T} \E\{ r^{(k)}_t \big\vert \mathcal{H}^{(k)}_{t-1} \} \right\} 
    \\
     & =  \E\bigg\{ \sum_{k=1}^K T \operatorname{val}(\boldsymbol{M}^{(k)})  - \sum_{t=1}^{T} \E\{\mu^{(k)} \boldsymbol{M}^{(k)} \nu_t^{(k)} \big\vert \mathcal{H}^{(k)}_{t-1} \} \\ 
     & \quad\quad\quad(\mathds{1}_{\{E_k\}} + \mathds{1}_{\{{E_k}^c\}})\bigg\}  
     \\
     & \leq \underbrace{\E\bigg\{ \sum_{k=1}^K \sum_{t=1}^T \operatorname{val}(\boldsymbol{M}^{(k)}) -  \E\{ \mu^{(k)} \boldsymbol{M}^{(k)} \nu^{(k)} \big\vert \mathcal{H}^{(k)}_{t-1} \}\mathds{1}_{\{E_k\}}\bigg\}}_{\text{(A) ineq. holds with } E_k }  \\ & \quad \quad +  \underbrace{KT \mathbb{P}( \bigcup_{k=1}^{K}{E_k}^c)}_{\text{(B) bounded by lemma \ref{confidenceball}}}.
 \end{aligned}
\end{equation*}
Here, the inequality is obtained by assumption \ref{regularity} (a).
Since the term (B) can be bounded as a constant by choosing $\delta = \frac{1}{KT}$, we turn to look at term (A). The term (A) can be viewed as the expected pseudo saddle-point regret when $E_k$ holds for all $k$, in this case, by the tower rule and the optimistic exploration:
\begin{equation*}
    \begin{aligned}
         & \E\bigg\{ \sum_{k=1}^K  T \operatorname{val}(\boldsymbol{M}^{(k)}) - \sum_{t=1}^{T} \E\{\mu^{(k)} \boldsymbol{M}^{(k)} \nu_t^{(k)} \big\vert \mathcal{H}^{(k)}_{t-1} \} \bigg\}\\
         = & \E\bigg\{ \sum_{k=1}^K  \sum_{t=1}^{T} \E \{ \operatorname{val}(\boldsymbol{M}^{(k)}) -\mu^{(k)} \boldsymbol{M}^{(k)} \nu_t^{(k)} \big\vert \mathcal{H}^{(k)}_{t-1} \} \bigg\} \\
        \leq & \E\bigg\{ \sum_{k=1}^K  \sum_{t=1}^{T} \E \{ \operatorname{val}(\widetilde{\boldsymbol{M}}^{(k)}) -\mu^{(k)} \boldsymbol{M}^{(k)} \nu_t^{(k)} \big\vert \mathcal{H}^{(k)}_{t-1} \} \bigg\} \\
        \leq & \E\left\{ \sum_{k=1}^K \sum_{t=1}^{T} \E\left[ \mu^{(k )\top} \widetilde{\boldsymbol{M} }^{(k)} \nu^{(k)} - \mu^{(k)} \boldsymbol{M}^{(k)} \nu^{(k)} \big\vert \mathcal{H}^{(k)}_{t-1} \right] \right\} .
    \end{aligned}
\end{equation*}
Therefore, when event $E_k$ holds for all $k$ with probability $1-\delta$, consider the term (A):
\begin{equation*}
     \begin{aligned}
      &\E\left\{ \sum_{k, t} \E\left[ \mu^{(k )\top} \widetilde{\boldsymbol{M} }^{(k)} \nu^{(k)} - \mu^{(k)} \boldsymbol{M}^{(k)} \nu^{(k)} \big\vert \mathcal{H}^{(k)}_{t-1} \right] \right\} 
     \\
      = &\E\left\{ \sum_{k, t}\E \left[  \mu^{(k )\top} (\widetilde{\boldsymbol{M} }^{(k)} - \boldsymbol{M}^{(k)}) \nu^{(k)} \big\vert \mathcal{H}^{(k)}_{t-1} \right] \right\}  
     \\
      = &\E\left\{  \sum_{k, t} \E \left[ \mu^{(k )\top} \underbrace{\sum_{s=1}^S (\tilde{\theta}_s^{(k)}- \theta^*_s) M^{(k)}_s}_{\text{matrix }\boldsymbol{h}( \tilde{\theta}^{(k)} - \theta^*, \mathcal{M}^{(k)} )}\nu^{(k)} \big\vert \mathcal{H}^{(k)}_{t-1} \right] \right\} .
     \end{aligned}
\end{equation*}

Here, each component of the matrix $\boldsymbol{h} (\tilde{\theta}^{(k)} - \theta^*, \mathcal{M}^{(k)})$ with subscripts $i, j$ is an inner product $\langle \tilde{\theta}^{(k)} - \theta^* , z^{(k)}_{i, j}\rangle$, and thus, 
\begin{equation*}
    \begin{aligned}
    & \E \left\{ \sum_{k, t} \E \left[ \sum_{i, j} \mu_i^{(k)} \langle \tilde{\theta}^{(k)} - \theta^*, z_{ij}^{(k)}  \rangle \nu_j^{(k)} \big\vert \mathcal{H}^{(k)}_{t-1} \right] \right\} \\
       = &\E \left\{ \sum_{k, t} \E \left[ \E \left( \langle \tilde{\theta}^{(k)} - \theta^*, z_{i_t, j_t}^{(k)}  \rangle  \big\vert \mathcal{H}^{(k)}_{t-1} \right) \right] \right\}\\
       \leq &\E \left\{ \sum_{k, t} \E \left[ \| \tilde{\theta}^{(k)} - \theta^*\|_{V_{k-1}} \| z_{i_t, j_t}^{(k)}  \|_{V_{k-1}^{-1}} \big\vert \mathcal{H}^{(k)}_{t-1}  \right] \right\} \\
       \leq  &\E\bigg\{ \sum_{k, t} \E\bigg[ (\| \theta^* - \hat{\theta}^{(k)}\|_{V_{k-1}} + \| \tilde{\theta}^{(k)} - \hat{\theta}^{(k)}\|_{V_{k-1}})  \\  & \quad \quad \quad \quad \| z_{i_t, j_t}^{(k)}  \|_{V_{k-1}^{-1}} \big\vert \mathcal{H}^{(k)}_{t-1} \bigg] \bigg\}  \\
      \leq & \E\left\{ \sum_{k, t} \E\left[ 2 \sqrt{\beta_k(\delta)} \| z_{t}^{(k)}  \|_{V_{k-1,t}^{-1}} \sqrt{\frac{\operatorname{det}(V_{k-1,t})}{\operatorname{det}(V_{k-1,0})}} \big\vert \mathcal{H}^{(k)}_{t-1} \right] \right\} \\
      \leq &\E\left\{ \sum_{k, t} \E\left[ 2 \kappa \sqrt{\beta_k(\delta)} \| z_{t}^{(k)}  \|_{V_{k-1,t}^{-1}} \big\vert \mathcal{H}^{(k)}_{t-1} \right] \right\} ,
    \end{aligned}
\end{equation*}
where we apply the Cauchy–Schwarz inequality for the first inequality; the triangular inequality for the second inequality; Lemma \ref{technicaltool1} for the third inequality; and Assumption \ref{regularity} (b) for the fourth inequality.

The pseudo saddle-point regret can be bounded up to now. With probability at least $1 - \delta$, the pseudo saddle-point regret, which is inside the expectation in term (A), satisfies:
 \begin{equation*}
     \begin{aligned}
       S\hat{\mathcal{R}}& \leq \sum_{k=1}^{K}  \sum_{t=1}^{T} 2\kappa  \sqrt{\beta_k(\delta)} \| z_{i_t, j_t}^{(k)} \|_{V_{k-1,t}^{-1}}.  
       \end{aligned}
\end{equation*}
According to Assumption \ref{regularity}, the regret per round is bounded by $2$, and since $\beta_K \leq \max\{1, \beta_k\}$,
\begin{equation*}
   S\hat{\mathcal{R}}  \leq \sum_{k= 1}^K 2\sqrt{\beta_K(\delta)} \sum_{t=1}^T\min\left(1, \|z_{i_t, j_t}^{(k)}\|_{V_{k-1}^{-1}} \right)  .
\end{equation*}
Using the quadratic mean inequality, we arrive at
\begin{equation*}
    S\hat{\mathcal{R}} \leq 2\sqrt{K T \beta_K(\delta) \kappa \sum_{k=1}^K \sum_{t=1}^T \min \left\{ 1,   \| z_{i_t, j_t}^{(k)}\|^2_{V_{k-1,t}^{-1}} \right\} } 
\end{equation*}
Applying Lemma \ref{boundingz} completes the proof of theorem \ref{pseudoregretofulinmat}. 
Finally, let $\delta = 1/KT$, we obtain:
\begin{equation*}
    S\mathcal{R} \leq \mathcal{O}(S \sqrt{KT} \ln(KTS)),
\end{equation*}
thus proving Corollary \ref{corollary}.

\subsection{Computational Issue}
While the following parameterized optimization problem is normally intractable:
$$
  \max_{\theta \in \mathcal{C}_k(\delta) \bigcap \mathcal{B}} \max_{\mu \in \Pi^1} \min_{\nu \in \Pi^2} \mu^{\top} (\sum_{s=1}^S \theta_s M_s^{(k)}) \nu,
$$
in some special case the optimistic strategy $\tilde{\mu}$ can be computed efficiently. For example, when $\mathcal{B}$ covers $\mathcal{C}_k(\delta)$, the total confidence set is just the original elliptical ball; i.e., $\mathcal{C}_k(\delta) = \mathcal{C}_k(\delta) \bigcap \mathcal{B}$, then the computation of $\tilde{\mu}$ and $\tilde{\theta}$ can be written as:
\begin{equation*}
     (\tilde{\mu}^{(k)}, \tilde{\theta}^{(k)}) = \argmax_{\theta \times \mu \in (\mathcal{C}_k(\delta) \times \Pi^1)} \min_{\nu \in \Pi^2} \mu^{\top} \left(\sum_{ s = 1 }^S \theta_s  M_s^{(k)}\right) \nu.
\end{equation*}
One can define the unit ball $\mathcal{B}_2 := \left\{x \in \mathbb{R}^{S}:\|x\|_{2} \leq 1\right\} $ and rewrite the elliptical ball: $\mathcal{C}_k(\delta) = \hat{\theta} +  \sqrt{\beta_k(\delta)} V_{k-1}^{-\frac{1}{2}} \mathcal{B}_2$. And since by Von Neumann minimax theorem the value is convex and concave in  $(\mathcal{C}_k(\delta) \times \Pi^1)$ and $\Pi^2$, we rewrite the objective as:
\begin{equation*}
\begin{aligned}
       (\tilde{\mu}^{(k)}, \tilde{\theta}^{(k)})&  =  \min_{\nu \in \Pi^2}  \argmax_{\theta \times \mu \in (\mathcal{C}_k(\delta) \times \Pi^1x)}\mu^{\top} \left[\langle \theta,  z_{i,j}^{(k)} \rangle \right] \nu \\
     \tilde{\mu}^{(k)}   =   \min_{\nu \in \Pi^2} & \argmax_{ \mu \in  \times \Pi^1} \mu^{\top} \left[\langle \hat{\theta},  z_{i,j}^{(k)} \rangle + \beta^{\frac{1}{2}}_k(\delta) \|z_{i,j}^{(k)}\|_{V_{k-1}^{-1}} \right] \nu,
\end{aligned}
\end{equation*}
where $[\cdot]$ represents the $n_1 \times n_2$ matrix formed by $i,j$-th entries. In doing so, we compute the optimistic estimate of every game entry.

%%%%%%%%%%%%%%%%%%%%%%%%%%%%%%%%%%%%%%%%%%%%%%%%%%%%%%%%%%%%%%%%%%%%%%%%%%%%%
\subsection{Adversarial Algorithm} \label{exp3framework}
In the viewpoint of a player who cannot observe what action her opponent takes at each round, 
the interaction per episode can be formulated as a adversarial multi-armed bandit problem. The arm set is the set of rows $\{1, \ldots, n_1\}$, and we shall assume that the adversary selects the worst-case column sequences. This assumption is equivalent to selecting a sequence of loss vector. Hence, denote $e_{it}$ as the one-hot basis vector, the row player receives:
\begin{equation*}
    r^{(k)}_t = e_{i_t}^{\top} \boldsymbol{M}^{(k)}_{j_t},
\end{equation*}
which is essentially an adversarial MAB problem. \texttt{Exp3} is a common approach for this type of problem.
One version of the \texttt{Exp3} algorithm outputs the policy based on the cumulative reward estimates of each action.
\begin{equation*}
    \mu^{(k)}_{t,i} = \alpha_t \frac{1}{n_1} + (1- \alpha_t) \frac{\exp(\gamma_t \sum_{t} \hat{r}^{(k)}_{t,i})}{ \sum_{i=1}^{n_1}\exp(\gamma_t \sum_{t} \hat{r}^{(k)}_{t,i})},
\end{equation*}
where $\alpha_t$ is the weight for uniform exploration, $\gamma_t$ is the learning rate, and $\hat{r}^{(k)}_{t,i}$ is obtained through the importance-sampling estimator:
\begin{equation*}
    \hat{r}^{(k)}_{t,i} = \mathds{1}_{\{i_t = i\}} \frac{r^{(k)}_t}{\mu^{(k)}_{t,i}}.
\end{equation*}
It has been shown that when $\alpha_t$ and $\gamma_t$ are tuned properly, the algorithm leads to a $\tilde{\mathcal{O}}(\sqrt{n_1 T})$ best-response regret bound per episode, thus an \texttt{Exp3} agent suffers totally  $B\mathcal{R} \leq \tilde{\mathcal{O}}(K\sqrt{n_1 T})$.

%%%%%%%%%%%%%%%%%%%%%%%%%%%%%%%%%%%%%%%%%%%%%%%%%%%%%%%%%%%%%%%%%%%%%%%%%%%%%%%%
\section{CASE STUDY}\label{casestudy}
\begin{figure}[htbp]
    \centering
    \includegraphics[width = .75\textwidth]{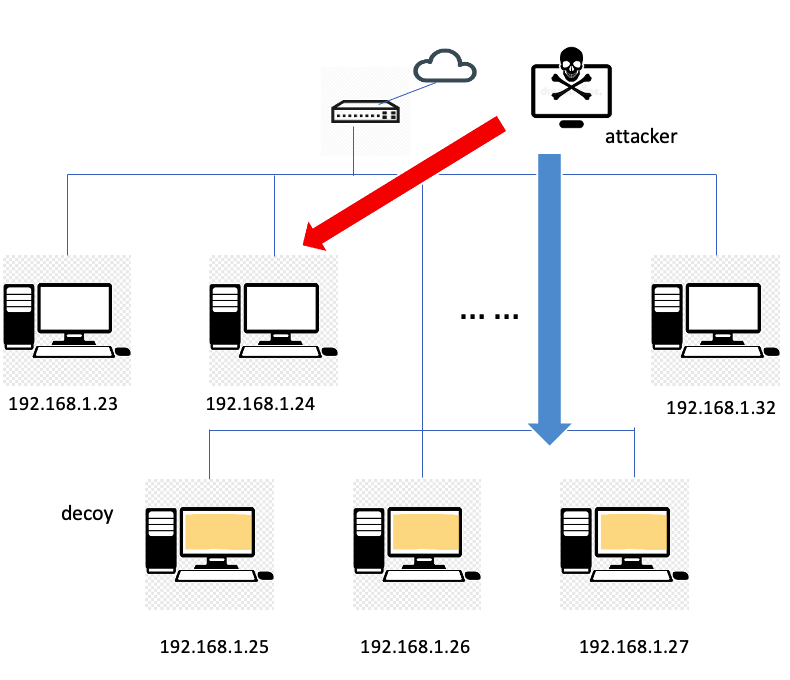}
    \caption{The dynamic honeypot allocation (DHA) game: The attacker moves in the network and selects node to attack while the defender aims to bait with the honeybots.}
    \label{honepotallocation}
\end{figure}

Game theory has played an important role in modeling the strategic interaction between a system defender and an adversary \cite{pawlickgame,pawlick2019game,manshaei2013game}. In many security games, the attack model and the game matrix are prescribed by the designer, who aims to protect the system from a known class of attacks. In many security applications, the capabilities of the attackers are not fully known to the defender. Hence, it is essential to develop mechanisms for the defender to update her strategies online. In this case study,
we consider a version of Dynamic Honeypot Allocation (DHA) game. In this game, a defender places decoys to protect network resources whereas some periphery attackers aim to capture these resources. Thus, the action pairs of attacking or placing decoys in each of the nodes construct a matrix game, with unknown payoffs. The environment contains $S$ experts, an agent for central allocation and an attacker, and an underlying time-varying matrix game. Fig. \ref{honepotallocation} illustrates one round of play, the attacker and the defender simultaneously select a subset of nodes to attack and place decoys, the outcome of the play is encoded in the underlying matrix game.

\subsubsection{Experimental Set Up}
Suppose that both players have $10$ strategies, $n_1 = n_2 = 10$. Let the defender interact with the opponent for $15$ episodes, $200$ rounds per episode.  
We fix a true sampled parameter $\theta^* \sim \mathcal{N}(0.5 \boldsymbol{1}_S, I_{S\times S})$, and sample $10$ $10 \times 10$ matrices for every episode from uniform distribution $\mathcal{U}[0, 1]$, together yielding the true game matrices $\boldsymbol{M}^{(k)}$. 
The interaction outcome is the entry of $\boldsymbol{M}^{(k)}$ plus an i.i.d. noise $\eta^{(k)}_t$ sampled from Gaussian $\mathcal{N}(0,0.5)$. 

\subsubsection{Methodology and Results}
We show that the regret of \texttt{OFULinMat} is converging in a sublinear speed, which is much faster than \texttt{Exp3}. In running the former, we set $\lambda = 0.1$, $\delta = 3 \times 10^{-3}$, and $B = 3$. In running the latter, we set the parameter $\alpha_t = \min(1, \sqrt{n_1 \ln n_1 / t})$ and $\gamma_t = \sqrt{2\ln n_1 / n_1 t}$, and reset the cumulative estimates for every episode. We let defender use these two algorithms in parallel to play in the same game, against an omniscient attacker who always play saddle-point strategies, and report the pseudo saddle-point regret for both players. We also report the parameter estimation process of \texttt{OFULinMat}. 

\begin{figure}[htbp]
    \centering
    \includegraphics[width = .8\textwidth]{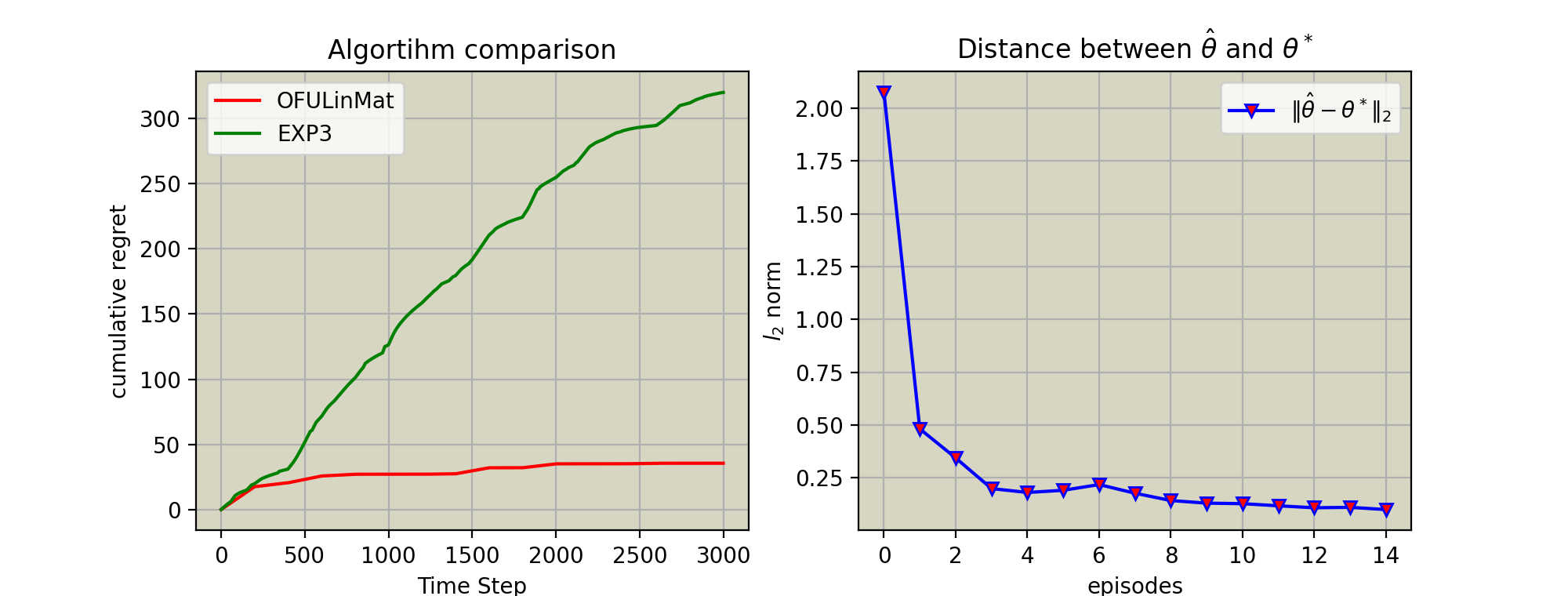}
    \caption{The figure shows that \texttt{OFULinMat} outperforms \texttt{Exp3}, meanwhile, the estimated parameter $\hat{\theta}$ also converges to the true parameter $\theta^*$.}
    \label{resultsfigure}
\end{figure}

The results are shown in Fig. \ref{resultsfigure}. It is clear that for \texttt{OFULinMat}, as the estimated parameter becomes more accurate, the regret per round becomes smaller and the cumulative regret curve becomes flatter. Taking advantage of additional expert information, its performance is much better than a naive \texttt{Exp3} agent. 

%%%%%%%%%%%%%%%%%%%%%%%%%%%%%%%%%%%%%%%%%%%%%%%%%%%%%%%%%%%%%%%%%%%%%%%%%%%%%%%%
\section{CONCLUSIONS AND FUTURE WORKS} \label{conclusion}

\subsection{Conclusions}
We have proposed an episodic online learning framework for a time-varying zero-sum game environment with expert game models. The learners do not know the entries of the game matrix and yet they have imperfect observations of the outcomes of their play. The proposed \texttt{OFULinMat} algorithm has addressed the learning problem by integrating the parameter estimation phase and optimistic exploration phase during the play. We have shown that our algorithm is provably efficient by establishing a sublinear upper bound on the saddle-point regret, under the model linearity assumption. Comparing it to the classical adversarial multi-arm bandit algorithm in the case study of dynamic honeypot allocation game, we have seen  that additional expert information can significantly boost the efficiency of the learning process.

\subsection{Future Works}
There are many future research directions related to the proposed framework. E.g., from a computational perspective, we can show whether Thompson sampling will be provably efficient in exploring the weighting coefficients, given that its implementation is often simpler than an algorithm with OFU principles. 
When the matrix becomes large and sparse, more efficient planning and estimation techniques need to be incorporated. Another potential direction lies in the model linearity assumption. It is possible to explore a richer class of parametric or non-parametric functions for game estimation and adapt the model ensembles during the learning process. The proposed framework could also inspire the meta-learning in games, which enables agents to quickly adapt to new tasks.

\bibliographystyle{unsrt}
\bibliography{ref}

\end{document}